\documentclass[10pt]{article}
\usepackage[margin=1.25in]{geometry}
\usepackage{graphicx}
\usepackage{amsmath}
\usepackage{amssymb}
\usepackage{authblk}
\usepackage[numbers]{natbib}

\newcommand*{\citen}[1]{%
  \begingroup
    \romannumeral-`\x 
    \setcitestyle{numbers}%
    \cite{#1}%
  \endgroup   
}

\bibpunct{}{}{,}{s}{}{,}

\begin{document}

\title{An Impedance-Modulated Code-Division Microwave SQUID Multiplexer}

\author[1]{C. Yu\footnote{\texttt{cyndiayu[at]stanford[dot]edu}}}
\author[1]{A. Ames}
\author[1]{S. Chaudhuri}
\author[1]{C. Dawson}
\author[1,2]{K.D. Irwin}
\author[1]{S.E. Kuenstner}
\author[2]{D. Li}
\author[1]{C.J. Titus}
\affil[1]{Department of Physics, Stanford University, Stanford, CA 94035, USA}
\affil[2]{SLAC National Accelerator Laboratory, Menlo Park, CA 94025, USA}
\date{}
\maketitle

\begin{abstract}

Large arrays of cryogenic detectors, including transition-edge sensors (TESs) or magnetic micro-calorimeters (MMCs), are needed for future experiments across a wide range of applications. Complexities in integration and cryogenic wiring have driven efforts to develop cryogenic readout technologies with large multiplexing factors while maintaining minimal readout noise. One such example is the microwave SQUID multiplexer ($\mu$mux), which couples an incoming TES or magnetic calorimeter signal to a unique GHz-frequency resonance that is modulated in frequency. Here, we present a hybrid scheme combining the microwave SQUID multiplexer with code division multiplexing: the impedance-modulated code-division multiplexer (Z-CDM), which may enable an order of magnitude increase in multiplexing factor particularly for low-bandwidth signal applications.


\end{abstract}

\section{Introduction}

Transition-edge sensors (TESs) are widely used in sub-millimeter astronomy, x-ray astrophysics, and x-ray spectroscopy at light sources. 
TESs provide background-limited sensitivity in the submillimeter, and both high spectral resolution and high efficiency for x-ray measurements [\citen{irwin95,ullom15,porter05,s4tech}]. 
Magnetic microcalorimeters (MMCs) [\citen{wegner2018microwave}] have shown promise for use in calorimetric applications, including x-ray astrophysics and spectroscopy [\citen{bandler04}]. 
As these systems are scaled to larger arrays, cryogenic multiplexing (MUX) is required to reduce the number of wires that run from the cryogenic temperature stage (typically 50 mK - 300 mK) to room temperature. 

Cryogenic multiplexing systems have now been developed and deployed at megahertz frequencies using time-division multiplexing (TDM) [\citen{chervenak99}], frequency-division multiplexing (FDM) [\citen{dobbs09}], and Walsh code-division multiplexing (CDM) [\citen{cdm10,morgan16}], and at gigahertz frequencies using microwave SQUID multiplexing ($\mu$mux) in both transition-edge sensors [\citen{irwin04, mates08}] and magnetic microcalorimeters [\citen{kempf2017demonstration}]. 
$\mu$mux has GHz readout bandwidth but uses that bandwidth inefficiently, achieving MUX factors of $\sim$4000. 
TDM, FDM, and CDM have higher Shannon efficiency, but only MHz available bandwidth, enabling MUX factors of $\sim$100 [\citen{irwin09}]. 

We propose the impedance-modulated code-division multiplexer (Z-CDM), an implementation of Walsh code-division multiplexing in GHz resonators, which has the potential for both high efficiency and GHz bandwidth, thus enabling MUX factors of 10,000 -- 100,000 per coaxial cable pair.
 Z-CDM uses multiple rf SQUIDs coupled to each resonator, with polarity modulation of the reactive load impedance that each SQUID presents to the resonator.
 Each microwave resonator can accomodate multiple polarity-modulated input signals, allowing for a large improvement in multiplexing factor across a typical 4-8GHz bandwidth.
 For simplicity, we refer specifically to coupling to TESs throughout the text, but this readout technology is also applicable to MMCs.

In Sec.~\ref{sec:mumux} we describe factors limiting the MUX factor of microwave SQUID multiplexers.
 We then present the impedance-modulated code division multiplexer and its operation in Sec.~\ref{sec:Z-CDM}, and discuss details of its implementation and consider non-idealities in Sec.~\ref{sec:implementation}. 

\section{Limitations of Microwave SQUID Multiplexers ($\mu$mux)}\label{sec:mumux}

\begin{figure*} [t]
    \centering
    \includegraphics[width=0.8\linewidth]{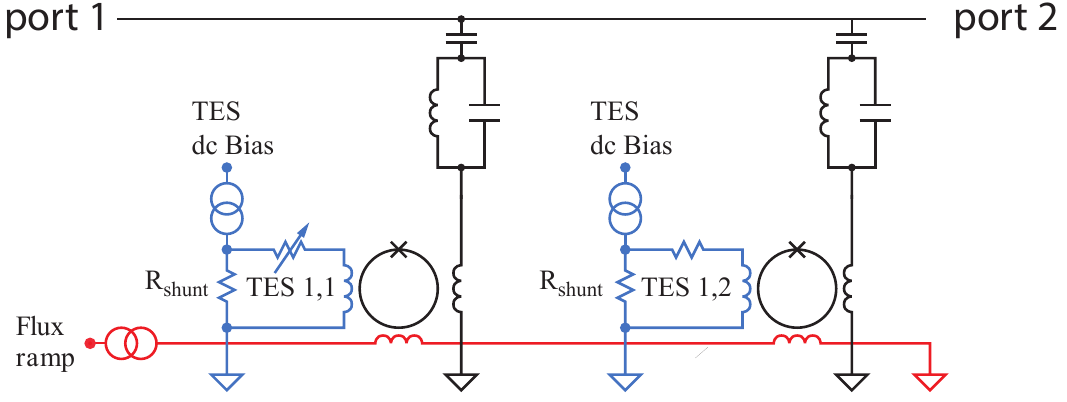}
    \caption{(color online) A schematic of the microwave SQUID multiplexer ($\mu$mux). Two pixels are shown here, though in practice thousands of resonators may be placed on the same microwave feedline. A common dc TES bias current (blue) passes through the shunt resistors $R_{\rm sh}$ and the TES detectors such that the TESs are voltage biased into electrothermal feedback. The current in the TES applies a flux to a dissipationless rf SQUID, which is in turn coupled to the microwave resonator. Each microwave resonator is tuned to a unique frequency, and the modulation of its effective inductance from the SQUID input causes its resonance frequency to change. A comb of excitation frequencies is incident at the input port (port 1) to the microwave feedline, while the transmitted signal carrying the status of each TES-coupled resonator is carried out to the amplifier and the warm readout electronics (port 2). A common flux ramp (red) is applied to all the SQUIDs in order to linearize them without the need for individual feedback lines.}
    \label{fig:umux}
\end{figure*}

The microwave SQUID multiplexer consists of an array of high-$Q$ microwave resonators with unique resonance frequencies, each inductively coupled to a dissipationless rf SQUID (an unshunted Josephson junction in a superconducting inductive loop) [\citen{irwin04,mates08}]. 
A schematic is shown in Figure~\ref{fig:umux}. 
Each rf SQUID is also inductively coupled to the current flowing through a single TES. The input TES signal modulates the effective inductive load $\Delta L_{\rm eff}$ that the rf SQUID presents to the resonator, shifting its resonance frequency. 
All of the resonators are coupled to a single feedline. 
A comb of excitation frequencies tuned to each resonance is incident at the input port to the microwave feedline, and the transmitted signal imprinted with the status of each TES-coupled resonator is carried out to the amplifier and the warm readout electronics. 

In this way, arrays of several thousand TESs may be read out with single pair of coaxial cables. 
The microwave SQUID response is linearized via the application of a common flux ramp drive to all SQUIDs. 
The flux ramp is a sawtooth with an integer number of flux quanta $\Phi_0$ in amplitude; thus the detector signal may be measured as a phase shift in the periodic SQUID response, which is linear in the TES input signal [\citen{mates08}].

The inductance that each rf SQUID presents to its resonance is a function of the flux $\Phi$ applied to the rf SQUID and given by: 

\begin{align}\label{eq:umux_leff}
\Delta L_{\rm eff}(\phi) &= - \frac{M_c^2}{L_S}\frac{\lambda\cos\phi}{1 + \lambda\cos\phi}\nonumber\\
&= -\frac{M_c^2}{L_S}\left(\lambda\cos\phi - \lambda^2\cos^2\phi + \cdots\right) \approx -\frac{M_c^2}{L_S}\lambda\cos\phi
\end{align}

\noindent
where $L_S$ is the self-inductance of the SQUID loop, $M_c$ is the mutual inductance between the SQUID and the resonator, and the dimensionless parameter $\lambda \equiv L_S/L_{J0}$.
 Here $L_{J0}=\Phi_0/(2\pi I_c)$  is the Josephson inductance of the junction in the SQUID, where $\Phi_0$ is the flux quantum and $I_c$ is the junction critical current. 
In typical $\mu$mux designs, $\lambda \sim 0.3$ to ensure non-hysteretic operation, although in practice it may be made arbitrarily small.
 On the right hand side of Equation \ref{eq:umux_leff}, we expand the inductance for small $\lambda$, and see that in this limit, inductance modulation is cosinusoidal.

The output bandwidth of $\mu$mux is often limited to one octave, typically 4-8 GHz  due to commercial availability of rf components and to avoid interference from harmonics of the resonator fundamental frequencies. 
The resonator spacing then determines the MUX factor: for instance, with a 1 MHz spacing between resonances, as many as 4000 resonators can be coupled to TESs and read out on a single pair of coaxial cables. 

In many TES applications in submillimeter, CMB, and x-ray astronomy, the required signal bandwidth in each TES is very small ($\sim$100 Hz -- 1 kHz), in which case, the achievable multiplexing factor is constrained by fabrication limitations on quality factor $Q$ of the resonator and random variations in resonator frequency placement $\Delta f_{\rm res}$. 
Typical linewidths achieved in fabrication are $\sim$100 kHz.

In order to avoid problematic crosstalk, resonances must typically be separated by 5-10 linewidths, resulting in $\Delta f_{\rm res}\approx 500$ kHz -- 1 MHz for these low-bandwidth applications. 
In principle, the small TES bandwidth in some applications would allow resonances to be placed $\sim1$ kHz apart, enabling MUX factors $> 10^6$. 
However, this would require line placement precision of $\Delta f_{\rm res}\sim$ kHz and $Q > 10^7$, instead of more typical achieved values of $\Delta f_{\rm res}\sim 1$ MHz and $Q \sim 10^5$.

Recently, progress has been made towards bringing microwave SQUID multiplexing systems to maturity for a variety of applications including particle physics [\citen{kempf2017demonstration, holmes18}], x-ray and $\gamma$-ray spectroscopy [\citen{mates17}], x-ray astrophysics [\citen{bennett19, lynx_concept}], and cosmology [\citen{s4tech, dober17,  so18, ari19, dober_ltd}]. 
In order to achieve their desired multiplexing factors given current fabrication techniques, designs relaxing the resonator placement requirements would ease implementation for these applications over the current $\mu$mux design. 
Applications particularly in sub-mm and x-ray astronomy benefit from achieving higher multiplexing factors than currently available with the present $\mu$mux design. 
Z-CDM enables both these relaxed placement requirements and these higher MUX factors. 

\section{Impedance-Modulated Code Division Multiplexer (Z-CDM)}\label{sec:Z-CDM}

\begin{figure*} [t]
    \centering
    \includegraphics[width=0.8\linewidth]{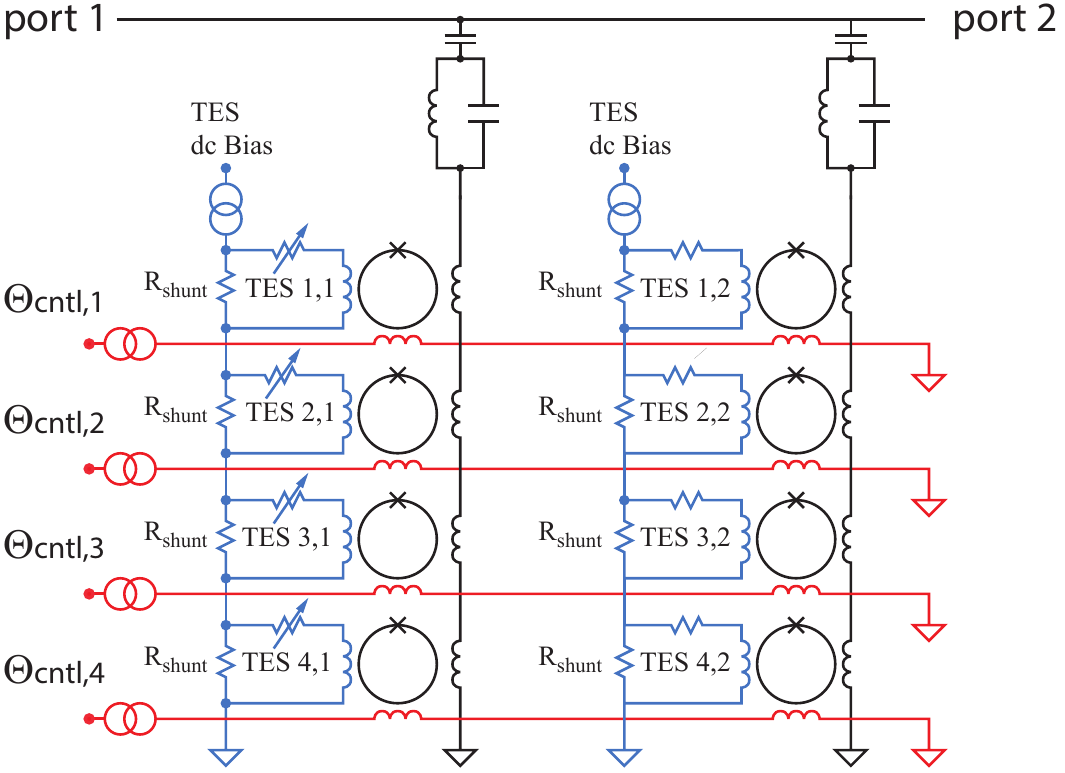}
    \caption{(color online) A circuit diagram of the impedance-modulated code division multiplexer (Z-CDM) with two resonances shown, each coupled to four SQUIDs and TESs for an eight-pixel implementation. The control lines (red) provide the flux ramp and the $\pi$ offset switching to rows of resonances, modulating the inputs from each TES (blue) coupled to a resonance in a Walsh pattern. Thus, relative to Figure~\ref{fig:umux} this represents a 4-fold increase in multiplexing factor with no increased requirements on resonator quality factor or spacing. }
    \label{fig:zcdm_2res}
\end{figure*}

The $\mu$mux circuit may be extended by code division multiplexing many TES detectors into each microwave resonator, allowing high MUX factors even with relatively low $Q$ and with significant errors in resonator line frequency placement $\Delta f_{\rm res}$. 

In existing flux-summed CDM ($\Phi$-CDM) multiplexers at MHz frequencies,  the signal from many TESs are inductively summed, with many input coils connecting to each SQUID [\citen{cdm10,morgan16}].
 Each of the SQUIDs is coupled to the input coils in different polarities, encoding a Walsh matrix. The signal from each TES can be demultiplexed at room temperature without degradation.
 In Z-CDM, the code division is instead done by summing the reactive impedance $Z = i \omega L$ from multiple rf SQUIDs into a single resonator, and then modulating the sign of the impedance of each rf SQUID.

\begin{figure*} [t]
    \centering
    \includegraphics[width=0.8\linewidth]{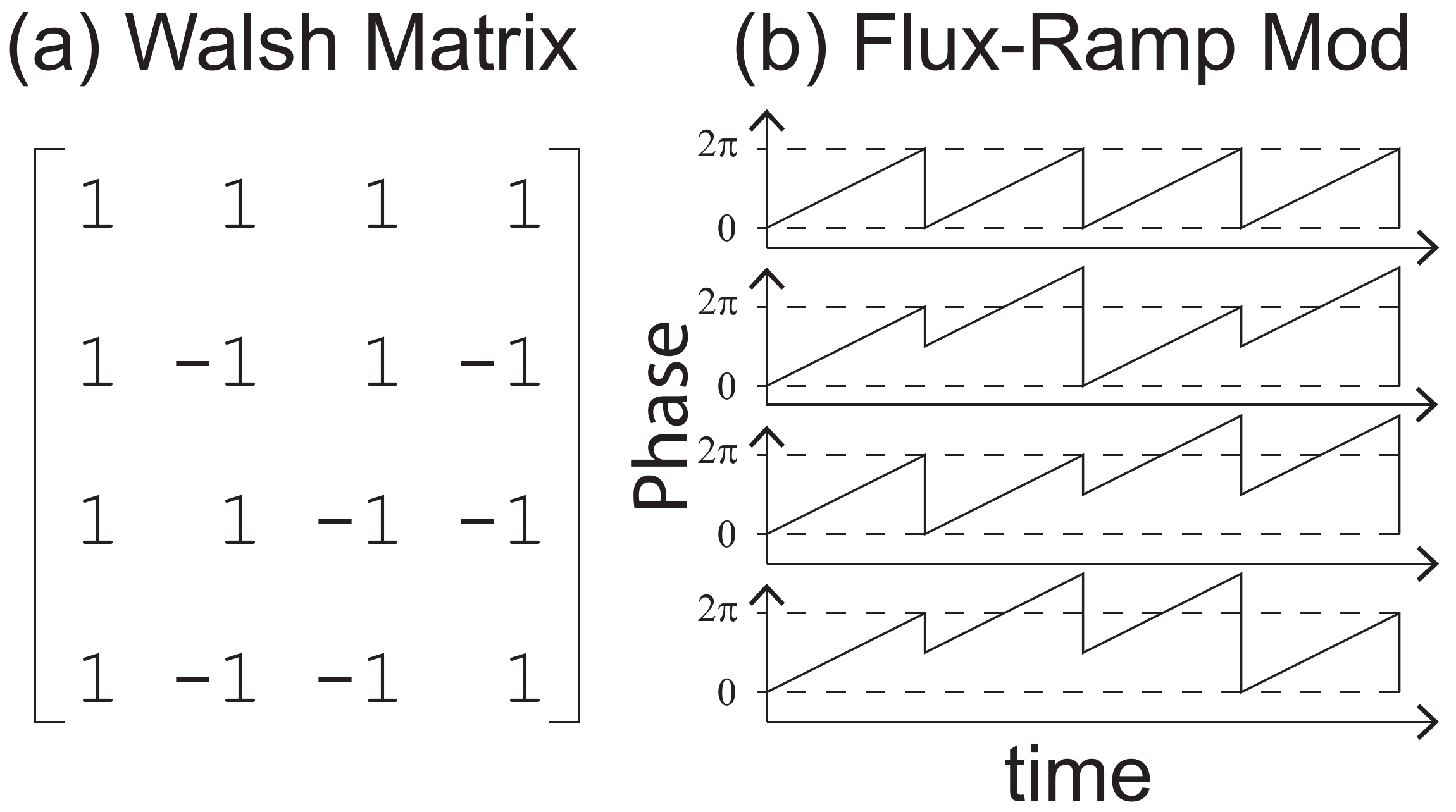}
    \caption{(a) An example Walsh code and (b) accompanying flux ramp waveform for $N_{\rm res}=4$ TESs coupled to each resonance. Each of the control lines addresses a row of TESs and may be shared across resonances as in Figure~\ref{fig:zcdm_2res}. Since we must fully sample a period of the SQUID curve in order to extract its lowest harmonic, we may treat a Walsh frame as a single flux ramp reset and modulate polarities via a DC offset in the flux ramp sawtooth. The $+1$ entries of the Walsh code have a flux ramp that sweeps from 0 to $\Phi_0$ (0 to $2\pi$ in phase), while the $-1$ entries have an additional $\pi$ phase offset and thus sweep from $\pi$ to $3\pi$ in phase.}
    \label{fig:modulation}
\end{figure*}

In this approach (shown in Figure~\ref{fig:zcdm_2res} for the example of $N_{\rm CDM}=4$ pixels coupled to each resonator), $N_{\rm CDM}$ common control lines provide a flux ramp modulation to the rf SQUIDs.
 If there are $N_{\rm res}$ resonators feeding into each output channel, this results in a MUX factor of $N_{\rm MUX} = N_{\rm CDM} \times N_{\rm res}$.
 This flux-ramp modulation is superimposed with a time-varying signal that modulates the polarity of the inductance shift caused by the resonator. 
 This modulation is provided by an additional $\Phi_0/2$ of flux, corresponding to an applied SQUID phase change of $\Delta\phi = \pi$, to modulate the polarity with which the SQUID couples to the resonator.
 This modulation pattern is shown in Figure~\ref{fig:modulation}, again for the example of $N_{\rm CDM}=4$. The $\pi$ phase offset is applied in an orthogonal polarity code, typically a Walsh matrix, and may subsequently be inverted in order to reconstruct the independent TES input signals. 

From Equation \ref{eq:umux_leff}, we see that a phase shift of $\Delta\phi = \pi$ is effectively a modulation of the sign of the polarity of $L_{\rm eff}$, and this approximation holds well for our small $\lambda$ limit. 

The modulation of the effective impedance of each resonance with $N_{\rm CDM}$ pixels presents a hybrid design combining the high bandwidth of $\mu$mux with the relatively high efficiency of MHz multiplexing techniques, including TDM, FDM, and CDM.
 This combination can enable a dramatic increase of 1--2 orders of magnitude in MUX factor over $\mu$mux in some applications.
 Additionally, since the TES input is coupled separately from the flux ramp lines which apply the offsets to the SQUIDs, this implementation allows for fast switching without the backaction in current-steering CDM that has limited its application [\citen{cdm10}].

For a desired multiplexing factor, this technique alleviates pressures on resonator $Q$ and frequency spacing by allowing resonances to be spaced more coarsely, while achieving high MUX factors. 
Where previously in $\mu$mux the target resonator spacing is set to approximately 10 bandwidths apart, for $N_{\rm CDM}$ SQUIDs per resonance the requirement on spacing becomes relaxed to $10 \times N_{\rm CDM}$ bandwidths apart. 
As resonator frequency placement has been a source of difficulty for current generations of $\mu$mux designs, this presents a promising path to achieving the multiplexing factors of several thousand envisioned. 

\section{Implementation and Performance Details}\label{sec:implementation}

Even using the same resonator spacing as in $\mu$mux of about 1MHz, the addition of multiple SQUIDs per line may increase the multiplexing factor by 1--2 orders of magnitude, allowing for 10,000 -- 100,000 TESs to be read out on a single pair of coaxial cables and a relatively small number of low-frequency modulation wires. 
The limit on the number of SQUIDs that may be coupled to a resonator $N_{\rm CDM}$ depends on a number of design choices that vary the gains to be achieved with this technique.

The combined slew of the SQUIDs must be smaller than the spacing between resonances in order to avoid resonator collisions as they modulate. 
For a given resonator spacing, this ultimately sets the limit on how many SQUIDs may be coupled to a resonator before encountering collisions. 
However, for the small-signal limit in which applications such as sub-mm astronomy operate, resonator spacings are wide compared with the signal and this does not impose a particularly stringent requirement on $N_{\rm CDM}$. 
In fact, an advantage of Z-CDM is its relaxation of resonator spacing requirements, which have traditionally been difficult to achieve in fabrication, in favor of more widely spaced resonances with multiple SQUIDs coupled to each resonator. 
The particular details of the lithography process are critical to assessing the accuracy with which resonances may be placed both in absolute frequency space and relative to each other; thus, the gains to be achieved with Z-CDM are extremely system dependent. 
Lithographed elements may additionally suffer from direct inductive crosstalk between SQUIDs for tightly packed layouts. 
However, this physical pickup has been found to be subdominant to other crosstalk mechanisms, particularly if neighboring resonators in frequency space are well-separated in their physical placement [\citen{mates19}].

In quarter-wave resonator designs as has been currently implemented for $\mu$mux, the number of SQUIDs that may be placed on a single resonance is limited by the requirement that the SQUIDs stay near the current antinode (the voltage node) of the resonator. 
As additional SQUIDs are added, if the SQUIDs are too close to the current node, the full value of $\Delta L_{\rm eff}$ will not couple to the resonator, reducing the signal-to-noise ratio, and lossy dielectrics in the rf SQUIDs will couple to electric fields in the resonator, potentially reducing $Q$ of the resonator. 
It is additionally important to stay near the voltage node in order to mitigate against two-level systems (TLS) noise that arises from driving voltage fluctuations proportional to the electric field across the dielectrics. 
In the most recently fielded $\mu$mux designs [\citen{dober_ltd}], allowing the SQUID to deviate by no more than $\lambda/10$ from the current antinode in order to keep couplings within a factor of 2 of each other allows for $N_{\rm CDM} \sim 8$.
A lumped element resonator design could allow more SQUIDs to be coupled per resonator, since these designs would allow for smaller voltage variations and thus reduced TLS contribution.

In order to effectively decode the TES signals, the control lines must switch significantly faster than the rate of the TES input signal, and sample through at least one full flux ramp period.
 Once one or more flux-ramp periods are sampled, a digital filter is applied to extract only the fundament flux-ramp modulation signal and discard all harmonics, since harmonics do not cleanly polarity modulate with a $\pi$ phase shift.
 In the small $\lambda$ limit, all of the power is in the fundamental, so the degradation of signal-to-noise ratio from discarding higher harmonics is small. 
 Since we intentionally design for small $\lambda$ in order to operate in the non-hysteric regime, this effect is far subdominant to other sources of noise in the system. 
 As in $\mu$mux systems, the flux ramp rate must be at least fast enough to get above the $1/f$ knee of TLS noise. 
 
In addition to TLS considerations, the flux ramp reset rate must also be fast enough to allow for a sufficiently fast Walsh sampling rate such that all the SQUIDs are sampled and demodulated before the signal changes too much. 
 Particularly for fast signals such as x-ray pulses, a large change over the course of a full Walsh sampling period causes crosstalk since the finite sample time to modulate the polarity of each of the rows is not accounted for in the simplest possible matrix inversion. 
 However, this gain drift problem occurs in $\Phi$-CDM and has been successfully corrected for in x-ray pulse detection using in-frame linear time correction. 
 This technique interpolates a linear ramp over the course of the Walsh sampled frame, thereby markedly reducing the cross-talk during the rising edge of x-ray pulses [\citen{fowler12}].

In low-frequency $\Phi$-CDM, imperfect implementation and inversion of the Walsh matrix because of mutual inductance variation between SQUID coils can result in crosstalk.
 Similarly, shifts of phase differing slightly from $\Delta\phi=\pi$ in different pixels will result in crosstalk between pixels in Z-CDM.
 Like $\Phi$-CDM, Z-CDM can be corrected on demultiplexing to eliminate this source of crosstalk. 
 We discuss this at length below. 

Crosstalk in $\Phi$-CDM is dominated by small variations in inductive coupling between the summing transformers and coupling from the feedback coil to the input coil of non-addressed SQUID channels.
 The latter is inapplicable to $\mu$mux-like systems, while the former has an analogy as variations in the SQUID couplings that vary the $\pi$ offset away from being a perfect polarity modulation.
 
Consider, for example, the simple case of a $N_{\rm CDM} = 2$, with two TES flux signals (the product of the TES current and input mutual inductance) $\Phi_1$ and $\Phi_2$, which are taken to be constant during a multiplexing frame.
The flux-ramp current $I_{\rm ramp}$ is common to both SQUIDs, resulting in a flux-ramp phase shift of $\Phi_{\rm FR}=M_{0}I_{\rm ramp}=\omega_{\rm FR}t$, where $M_0$ is the mutual inductance of the coupling between the control lines and the SQUID and the resulting flux-ramp phase change has angular frequency $\omega_{\rm FR}$.
 The CDM polarity-modulation control current $I_{\rm CDM}$ applies an additional $\Delta\Phi=\pi$ modulation to implement polarity switching for a Walsh code. 
 From Equation \ref{eq:umux_leff}, the fundamental component of the inductance modulation $L_1$ seen by the resonator when zero polarity-modulation control current is applied to both TESs is the sum of the two inductances:

\begin{align}
L_1 &= -\frac{M_c^2\lambda}{L_s}\left[\cos\left(\omega_{\rm FR}t + \Phi_1\right) + \cos\left(\omega_{\rm FR}t + \Phi_2\right)\right]\nonumber\\
&=-\frac{M_c^2\lambda}{L_s}\left[\cos\omega_{\rm FR}t(\cos \Phi_1 + \cos \Phi_2) - \sin\omega_{\rm FR}t(\sin \Phi_1 + \sin \Phi_2)\right]\nonumber\\
&=-\frac{M_c^2\lambda}{L_s}\left[I_1 \cos\omega_{\rm FR}t + Q_1 \sin\omega_{\rm FR}t\right]
\label{eq:leff_tot_1}
\end{align}

\noindent where $I_1$ and $Q_1$ are the in-phase and quadrature components of $L_1$.

In the second frame of the Walsh code, a phase shift of $\Delta\phi = M_0 I_{\rm mod} \approx \pi$ is applied to only the second rf SQUID in order to apply a polarity modulation. 
However, small variations in control-line mutual inductance can result in deviations of $\Delta\phi = \pi + \epsilon$, resulting in crosstalk on demodulation if $\epsilon$ is not zero.
 The difference signal in the second frame is thus:

\begin{align}
L_2 &= -\frac{M_c^2 \lambda}{L_s}\left[\cos\left(\omega_{\rm FR}t + \Phi_1\right) + \cos\left(\omega_{\rm FR}t + \Phi_2 + \pi + \epsilon\right)\right]\nonumber\\
&= -\frac{M_c^2\lambda}{L_s}\left[\cos\omega_{\rm FR}t \left(\cos \Phi_1 - \cos\Phi_2  + \epsilon\sin\Phi_2 \right)\right.\nonumber\\
&\quad\quad\quad\quad - \left.\sin\omega_{\rm FR}t \left(\sin \Phi_1 - \sin \Phi_2- \epsilon\cos \Phi_2\right)\right]\nonumber\\
&=-\frac{M_c^2\lambda}{L_s}\left[I_2 \cos\omega_{\rm FR}t + Q_2 \sin\omega_{\rm FR}t\right]
\label{eq:leff_tot_2}
\end{align}

\noindent
where $I_2$ and $Q_2$ are the in-phase and quadrature components of $L_2$ as before, and we have expanded up to first order in small $\epsilon$.
 The Fourier series of the modulation signal extracts the amplitudes of the quadrature signals, making it possible to demultiplex and infer $\Phi_1$ and $\Phi_2$ from:

\begin{equation}
\label{eq:xtalk_matrix}
\left(\begin{array}{c}I_1 \\ I_2 \\ Q_1  \\ Q_2\end{array}\right) 
= \left(\begin{array}{cccc}
1 & 1 & 0 & 0\\
1 & -1 & 0 & \epsilon\\
0 & 0 & -1 & -1\\
0 & \epsilon & -1 & 1
\end{array}\right)
\left(\begin{array}{c}
\cos \Phi_1 \\ \cos \Phi_2 \\ \sin \Phi_1 \\ \sin \Phi_2
\end{array}\right)
\end{equation}

The matrix consists of two $2\times 2$ Walsh matrices (1 1; 1 -1) applied to each quadrature, with ($I_1, I_2$) and ($Q_1, Q_2$) as estimators of the input TES fluxes, which are statistically combined for optimal signal to noise ratio.
 A nonzero $\epsilon$ resulting from small variations in inductive coupling $M_0$ breaks the clean separation on demultiplexing and biases the estimate of the $I_2$ and $Q_2$ components, inducing crosstalk.
 These variations can be measured, and Equation \ref{eq:xtalk_matrix} may be inverted to recover a demultiplexed TES signal with crosstalk corrected for small $\epsilon$.
 The process may also be extended to higher orders in $\epsilon$.
 This process is in Z-CDM is similar to measuring the summing inductances in $\Phi$-CDM to eliminate that source of crosstalk. 
In principle, this effect may be perfectly corrected for arbitrarily large arrays and is expected to be subdominant to other crosstalk sources in the system. 

Finally, we mention that changes in mutual inductance will also result in small variations in flux-ramp frequency, which will result in a phase-dependent mis-measurement of the Fourier series-components.
 This effect is also encountered in $\mu$mux. 
 However, this mis-measurement polarity modulates with Walsh switching, so it is not a source of crosstalk at first order.

\section{Conclusion}\label{sec:conclusion}

The impedance-modulated code division multiplexer improves upon $\mu$mux by coupling each resonance to multiple TES inputs, allowing for higher multiplexing factors than achievable with conventional $\mu$mux designs.
 This advantage is particularly relevant for applications in the sub-mm and x-ray regime. For current $\mu$mux applications, if the multiplexing factor is held fixed, the combination of multiple TESs on each resonance relaxes the resonator $Q$ and frequency placement requirements, which allow for improved crosstalk and yield performance in existing systems. 
While the discussion here centers primarily on TES applications, we note that the design presented is applicable to magnetic micro-calorimeters (MMCs) as well.
This hybrid design combines the technological promise of $\mu$mux with the existing heritage of CDM, with potential applications across a wide range of signal frequencies.

\section*{Acknowledgements}
This work was supported in part by NASA under the Astrophysics Research and Analysis Program, grant \# NNX17AE65G. C. Yu was supported by the NSF Graduate Research Fellowship Program. 

\pagebreak

\end{document}